\documentclass[pra,10pt, twocolumn, floatfix, superscriptaddress]{revtex4-1}
\usepackage{amsmath} % need for subequations
\usepackage{graphicx} % need for figures
\usepackage{verbatim} % useful for program listings
\usepackage{color} % use if color is used in text
\usepackage{subfigure} % use for side-by-side figures
\usepackage{hyperref} % use for hypertext links, including those to external documents and
\usepackage{sidecap}
\usepackage{tikz}
\usepackage{amsmath,bm}
\usepackage{amsfonts}
\usepackage{mathtools}
\usepackage{epstopdf}
\usepackage{float}
\usepackage{physics}

\begin{document}

\title{Quantum knots in Bose--Einstein condensates 
created
%induced 
by counterdiabatic control}
\author{T. Ollikainen}
\email{tuomas.ollikainen@aalto.fi}
\affiliation{QCD Labs, COMP Centre of Excellence, Department of Applied Physics, Aalto University, P.O. Box 13500, FI-00076 Aalto, Finland}
\author{S. Masuda}
\affiliation{QCD Labs, COMP Centre of Excellence, Department of Applied Physics, Aalto University, P.O. Box 13500, FI-00076 Aalto, Finland}
\affiliation{College of Liberal Arts and Sciences, Tokyo Medical and Dental University, Ichikawa, Chiba 272-0827, Japan}
\author{M. M\"ott\"onen}
\affiliation{QCD Labs, COMP Centre of Excellence, Department of Applied Physics, Aalto University, P.O. Box 13500, FI-00076 Aalto, Finland}
\affiliation{University of Jyv\"askyl\"a, Department of Mathematical Information Technology, P.O. Box 35, FI-40014 University of Jyv\"askyl\"a, Finland}
\author{M. Nakahara}
\affiliation{QCD Labs, COMP Centre of Excellence, Department of Applied Physics, Aalto University, P.O. Box 13500, FI-00076 Aalto, Finland}
\affiliation{Department of Mathematics, Shanghai University, 99 Shangda Road, Shanghai, 200444, China}
\affiliation{Department of Physics, Kindai University, Higashi-Osaka, 57-8502, Japan}

\keywords{dilute Bose gas, Bose-Einstein condensation, topological defect, quantum knot}

\begin{abstract}
We theoretically study the creation of knot structures in the polar phase of spin-1 BECs using the counterdiabatic protocol in an unusual fashion. We provide an analytic solution to the evolution of the external magnetic field that is used to imprint the knots. As confirmed by our simulations using the full three-dimensional spin-1 Gross--Pitaevskii equation, our method allows for the precise control of the Hopf charge as well as the creation time of the knots. The knots with Hopf charge exceeding unity display multiple nested Hopf links.
\end{abstract}

\maketitle

\section{\label{sec:introduction}Introduction}

A knot, defined as a closed curve with possible links and crossings, is an important mathematical concept appearing in various branches of physics. Knots have been proposed as an early model for atoms~\cite{Thomson:1867}, stable configurations in electromagnetism~\cite{Arrayas:2017}, and as stable finite-energy solutions in three-dimensional classical field theory~\cite{Faddeev:1997}. They have been observed in various physical systems: in knotted vortex-lines in water~\cite{Kleckner:2013} and light~\cite{Leach:2005}, nematic liquid crystals~\cite{Tkalec:2011}, and DNA nanostructures~\cite{Han:2010}. In the context of quantum mechanics, knots were predicted and recently observed in the nematic vector field in spin-1 Bose--Einstein condensates (BECs)~\cite{Kawaguchi:2008,Hall:2016}. 

Topologically stable knots in continuous fields are nontrivial mappings from $S^3$ to $S^2$~\cite{Nakahara:2003}. They are characterized by the third homotopy group $\pi_3(S^2)\cong\mathbb{Z}$ and present an example of nonsingular topological defects. The topological invariant characterizing the knots is the integer-valued Hopf charge $Q$. It can also be referred to as the knot linking number, because the preimages of the points in $S^2$ constitute loops which are linked together exactly $Q$ times.

In addition to knots, there are numerous topological structures available in gaseous BECs with spin degree of freedom. Recent decades have shown predictions and observations of various types of vortices~\cite{Matthews:1999,Leanhardt:2003,Donadello:2014,Seo:2015,Borgh:2017}, solitons~\cite{Denschlag:2000,Strecker:2002,Khaykovich:2002}, monopoles~\cite{Stoof:2001,Savage:2003,Pietila:2009_2,Pietila:2009,Ray:2014,Ray:2015}, and skyrmions~\cite{Ruostekoski:2001,Leslie:2009,Choi:2012_2,Orlova:2016} in this exquisite system. Furthermore, the stability and dynamics of the defects are available for detailed exploration~\cite{Anderson:2001,Ruostekoski:2003_2,Ruostekoski:2003,Shin:2004,Choi:2012,Ollikainen:2014,Tiurev:2016,Ollikainen:2017_2}.

In the context of spin-1 BECs, it was recently shown that a so-called counterdiabatic (CD)~\cite{Demirplak:2003,Berry:2009} protocol can be used to accelerate the topological vortex creation and pumping processes, as well as to reduce the atom losses and unwanted spin transitions inevitably present in the topological vortex creation process~\cite{Masuda:2016,Ollikainen:2017}. In contrast, we use the CD protocol in an unusual way for the creation of a knot structure in the nematic vector field of the spin-1 BEC in the polar phase. In our scheme, the CD magnetic field is calculated in such a way that it induces a $\pi$-rotation on the nematic vector only along a predetermined ring in the condensate. The imperfect rotation elsewhere is utilized in our scheme for the creation of knots. 

We characterize the created knot structures in terms of the particle density distributions of different spin states, the associated Hopf charge, and the linked preimage rings. We investigate the effect of the finite knot creation time on these quantities as well as to the spin density distributions and show that the polar phase decays into the ferromagnetic phase for long knot creation times. Interestingly, we show that the radius of the ring that characterizes the core of the knot can be conveniently controlled with the parameters related to the CD protocol and choosing a short core radius leads to nested knots with high Hopf charges.

This paper is organized as follows. In Sec.~\ref{sec:theory}, we present the mean-field theory of spin-1 BECs, the topological considerations of the order parameter spaces together with the Hopf charge, and the utilized knot creation method using the CD magnetic fields. In Sec.~\ref{sec:results}, we present the numerical results on the creation of knots and describe the nontrivial topology related to cases with high Hopf charge, $Q>1$. Section~\ref{sec:conclusion} concludes the paper.

\section{\label{sec:theory}Theory}

\subsection{\label{sec:meanfield}Mean-field theory}

The mean-field order parameter of the spin-1 BEC can be written as $\Psi({\bf r},t)=\sqrt{n({\bf r},t)}e^{i\phi({\bf r},t)}\zeta({\bf r},t)$. Here, the $n$ is the particle density, $\phi$ is the scalar phase, and $\zeta=(\zeta_{+1},\zeta_0,\zeta_{-1})^T_\mathrm{Z}$ is the complex-valued three-component spinor with $\zeta^\dagger\zeta=1$. The subscript in the spinor components refers to the magnetic quantum number of the $z$-quantized spin states $\left\{ \left| +1 \right>,\left| 0 \right>,\left| -1 \right>\right\}$. %Vectors in the Cartesian basis are written without the subscript.

In the simulations, the condensate dynamics is solved within the mean-field approximation according to the Gross--Pitaevskii (GP) equation
\begin{align}
&i\hbar \partial_t\Psi({\bf r},t) = \left[-\frac{\hbar^2}{2m}\nabla^2 + V({\bf r}) + c_0\Psi^\dagger({\bf r},t)\Psi({\bf r},t) \right.\nonumber \\
&\left.\vphantom{\frac{\hbar^2}{2m}}+ c_2\Psi^\dagger({\bf r},t){\bf F}\Psi({\bf r},t)\cdot{\bf F}+g_F\mu_{\text{B}}{\bf B}({\bf r},t) \cdot {\bf F}\right]\Psi({\bf r},t),\label{eq:hamiltonian}
\end{align}
where we employ the external optical potential $V({\bf r})=m\left[\omega_\rho^2\left(x^2+y^2\right)+\omega_z^2z^2\right]/2$ and the external magnetic field ${\bf B}({\bf r},t)$. The Cartesian vector ${\bf F}=(F_x,F_y,F_z)$ is composed of the standard dimensionless spin-1 matrices. The coupling constants for the density and spin interactions are $c_0=4\pi\hbar^2(a_0+2a_2)/(3m)$ and $c_2=4\pi\hbar^2(a_2-a_0)/(3m)$~\cite{Ohmi:1998,Ho:1998}, respectively, where the $s$-wave scattering lengths for $^{87}$Rb are given by $a_0=5.387$~nm and $a_2=5.313$~nm~\cite{vanKempen:2002}, and the atomic mass by $m=1.443\times10^{-25}$~kg. Furthermore, $g_F=-1/2$ is the Land\'e $g$ factor for $^{87}$Rb, $\hbar$ is the reduced Planck's constant, and $\mu_{\text{B}}$ is the Bohr magneton. The number of atoms is set to $N=2.1\times10^5$, and the trapping frequencies to $\omega_\rho=2\pi\times124~\mathrm{Hz}$ and $\omega_z=2\pi\times248~\mathrm{Hz}$ throughout the simulations, corresponding to an oblate condensate. % with horizontal Thomas--Fermi radius being twice the vertical radius.

The knot structures are created in the polar-phase order parameter of the spin-1 BEC using spatially and temporally varying external magnetic fields. For $^{87}$Rb, the coupling constant $c_2$ is negative, implying ferromagnetic interactions in the absence of external magnetic fields. At low magnetic fields, the polar phase is dynamically unstable and decays into the ferromagnetic phase. However, the timescale for the decay due to this instability exceeds the knot creation time in the presence of magnetic field gradient~\cite{Ray:2015,Hall:2016}. 

\subsection{\label{sec:topology}Topological considerations}

Taking the Euler angles $\alpha$, $\beta$, and $\gamma$ as successive rotations about $z$, $y$, and $z$ axes, respectively, the general spinor in the polar phase becomes~\cite{Ho:1998}
\begin{align}
\zeta_\mathrm{P} &= \mathcal{U}(\alpha,\beta,\gamma)\begin{pmatrix}0\\1\\0\end{pmatrix}_\mathrm{Z}= \frac{1}{\sqrt{2}}\begin{pmatrix} -e^{-i\alpha}\sin\beta\\ \sqrt{2}\cos\beta\\ e^{i\alpha}\sin\beta\end{pmatrix}_\mathrm{Z}\nonumber\\
&=\frac{1}{\sqrt{2}}\begin{pmatrix} -d_x+i d_y \\ \sqrt{2}d_z \\ d_x+ id_y \end{pmatrix}_\mathrm{Z},\label{eq:polarop}
\end{align}
where $\mathcal{U}=e^{-i\alpha F_z}e^{-i\beta F_y}e^{-i\gamma F_z}$. In the last identity we have expressed the spinor using the real-valued unit vector $\hat{\bf d}=(d_x,d_y,d_z)^T=(\cos\alpha\sin\beta,\sin\alpha\sin\beta, \cos\beta)^T$, referred to as the nematic vector. It defines the direction of magnetic order in the condensate. Using this vector, we can express the order parameter in the Cartesian basis as $\Psi=\sqrt{n}e^{i\phi}\hat{\bf d}$. 

The order parameter space for the polar spin-1 BEC is $\mathcal{O}_\mathrm{P}=\left[\mathrm{U}(1)\times S^2\right]/\mathbb{Z}_2$~\cite{Zhou:2001}, where the $\mathrm{U}(1)$ symmetry is attributed to the scalar phase $\phi$ and the $S^2$ symmetry to the vector $\hat{\bf d}$. Furthermore, the order parameter is invariant under the simultaneous transformations $\hat{\bf d}\rightarrow-\hat{\bf d}$ and $\phi\rightarrow\phi+\pi$, giving rise to the division by $\mathbb{Z}_2$ in $\mathcal{O}_\mathrm{P}$.

The nontriviality of the third homotopy group of the polar order parameter, $\pi_3(\mathcal{O}_\mathrm{P})\cong\mathbb{Z}$, allows the existence of knot structures in this phase. The related topological invariant, the Hopf charge $Q$, is defined as~\cite{Faddeev:1997,Kawaguchi:2008}
\begin{equation}
Q=\frac{1}{16\pi^2}\int d{\bf r}\sum_{i,j,k} \epsilon_{ijk}\mathcal{F}_{ij}({\bf r})\mathcal{A}_k({\bf r}),\label{eq:hopf}
\end{equation}
where $\mathcal{F}_{ij}=\hat{\bf d}\cdot(\partial_i \hat{\bf d}\times \partial_j\hat{\bf d})$ and $\mathcal{A}_i$ is implicitly defined by $\mathcal{F}_{ij}=\partial_i\mathcal{A}_j - \partial_j\mathcal{A}_i$. We note that $\mathcal{A}_i$ can be defined up to a gauge $\mathcal{A}_i\rightarrow\mathcal{A}_i+\partial_i\eta$, where $\eta$ is a scalar function. For the sake of convenient integration in Eq.~(\ref{eq:hopf}), one may choose such a gauge that one of the components of $\mathcal{A}$ is zero. %In the knot structure, the preimage of each point in the order parameter manifold consists of closed loops in the compactified real space, which is homeomorphic to $S^3$. %Hence, the Hopf charge is sometimes referred to as the linking number of the knot.

\subsection{\label{sec:knots}Creation of knots using counterdiabatic control of magnetic field}

Previously, knots have been created in an initially nematicly $z$-polarized BEC by suddenly introducing a quadrupole magnetic field $b_\mathrm{q}(x\hat{\bf x}+y\hat{\bf y}-2z\hat{\bf z})$ in the middle of the condensate~\cite{Kawaguchi:2008,Hall:2016}. Here, $b_\mathrm{q}$ is the strength of the gradient magnetic field. In the following discussion, we utilize the scaled coordinate system $(x',y',z')=(x,y,2z)$ for convenience. The spin rotations leading to the knot configuration in Refs.~\cite{Kawaguchi:2008,Hall:2016} are induced by the linearly increasing Larmor angular frequency $\omega_\mathrm{L}(r')=g_F\mu_\mathrm{B}b_\mathrm{q}r'/\hbar$, where $r'=\sqrt{x'^2+y'^2+z'^2}$. Knots with $Q=1$ are generated by allowing the Larmor precession to continue for $T_\mathrm{L}=2\pi\hbar/(g_F\mu_\mathrm{B}b_\mathrm{q}R')$, where $R'$ is the effective extent of the condensate. Thus the nematic vector experiences a full $2\pi$ rotation at radius $R'$.

%Here, in contrast, we show that the same knot configuration can be created using a specific magnetic field control obtained from the CD scheme~\cite{Masuda:2016,Ollikainen:2017}. In general, the CD magnetic field for a spin-1 system in the presence of a changing magnetic field ${\bf B}({\bf r},t)$ can be calculated with~\cite{Berry:2009}

Here, in contrast, we show that the knot configuration can be created using a dynamic magnetic field control obtained from the CD scheme~\cite{Masuda:2016,Ollikainen:2017}. In the CD scheme, we first select the reference adiabatic dynamics of the spin degree of freedom corresponding to the instantaneous eigenstates of the Zeeman Hamiltonian $\mathcal{H}_\mathrm{Z}=g_F\mu_\mathrm{B}{\bf B}({\bf r},t)\cdot{\bf F}$. In general, the CD magnetic field for a spin-1 system in the presence of a changing magnetic field ${\bf B}({\bf r},t)$ can be calculated with~\cite{Berry:2009}

\begin{equation}
{\bf B}_\mathrm{CD}({\bf r},t)=\frac{\hbar{\bf B}({\bf r},t)\times\partial_t{\bf B}({\bf r},t)}{g_F\mu_\mathrm{B}|{\bf B}({\bf r},t)|^2}.\label{eq:cdfield}
\end{equation}

Our starting point is to design the CD field for the case in which the bias field is linearly inverted as ${\bf B}_\mathrm{bias}(t)=B_0(1-2t/T)\hat{\bf z}$, where $B_0$ is the initial bias field strength and $T$ is the inversion time while $b_\mathrm{q}$ is kept fixed. Hereafter, the time $T$ is referred to as the knot creation time. Furthermore, we employ the cylindrical coordinate system $(\rho,\varphi,z)$ below. 

\begin{figure}[t!]
\centering
\includegraphics[width=0.42\textwidth]{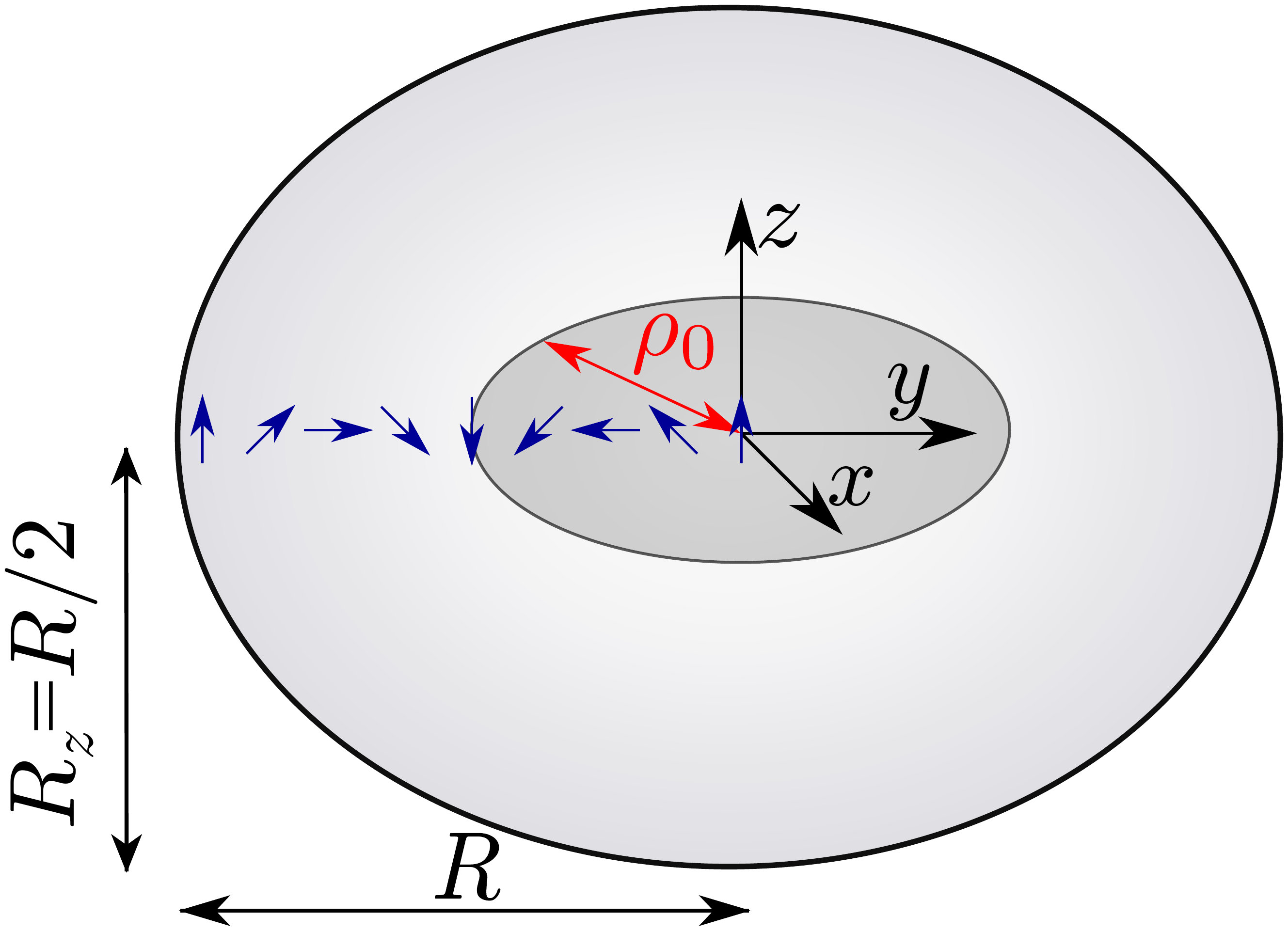}
\caption{(Color online) Orientation of the nematic vector (blue arrows) at certain points along $-y$-axis just after the quantum knot has been created. The $x$-component of the vector is zero along this axis. The origin is located in the middle of the condensate depicted by the shaded ellipsoid with radii $R$ and $R_z=R/2$, and $\rho_0$ is the radius of the circle along which the nematic vector $\hat{\bf d}$ rotates by $\pi$.}\label{fig:rotation}
\end{figure}

Application of the bias field inversion scheme directly into Eq.~(\ref{eq:cdfield}) leads to a CD field which rotates the nematic vector by $\pi$ everywhere. However, here we further set $z=0$ and $\rho=\rho_0$~\cite{Ollikainen:2017}. The thus employed magnetic field coincides with the original CD field only on the ring with radius $\rho_0$ in the $z=0$ plane, along which the nematic vector undergoes a $\pi$-rotation during the inversion of the bias field (see Fig.~\ref{fig:rotation}). This ring is referred to as the core of the knot structure. Indeed, a knot with $Q=1$ corresponds to the parameter choice $\rho_0=R/2$, where $R$ is the effective extent of the condensate in the $z=0$ plane. %{\color{red}With this choice, at radius $2\rho_0=R$ in the $z=0$ plane, the nematic vector experiences a full $2\pi$ rotation so that the order parameter assumes a constant value at the condensate boundary.}
Since the Larmor precession increases linearly as a function of distance from the origin, the nematic vector experiences a full $2\pi$ rotation at radius $2\rho_0=R$ so that the order parameter assumes a constant value at the condensate boundary. Along the $z$-axis the vector also retains its initial orientation. The nematic vector changes smoothly between these values. In practice, these rotations are induced by the brief pulse of magnetic field gradient near $t=T/2$, as is evident from the analytic form of the employed CD magnetic field shown below.

We further employ the unitary transformation introduced in Refs.~\cite{Masuda:2016,Ollikainen:2017} to obtain a CD field which can be experimentally implemented using a single pair of quadrupole coils. The transformation is given by $U(t)=e^{-i\alpha(t)F_z}$, where $\alpha(t)=\arctan\left[|{\bf B}_\mathrm{CD}|/\left(b_\mathrm{q}\rho\right)\right]$. As a result, the Zeeman part of the Hamiltonian for the unitary-transformed order parameter is rotated by $\alpha(t)$ and an additional time-dependent magnetic field is introduced along $z$. The resulting magnetic field giving rise to the knot structure is~\cite{Ollikainen:2017}

\begin{equation}
{\bf B}({\bf r},t) = b_\mathrm{q}^\mathrm{CD}(t)\left(x\hat{\bf x}+y\hat{\bf y}-2z\hat{\bf z}\right)+ B_0^\mathrm{CD}(t)\hat{\bf z},\label{eq:cdfield1}
\end{equation}
where 
\begin{equation}b_\mathrm{q}^\mathrm{CD}(t)=b_\mathrm{q}\sqrt{1+\left\{\frac{2\hbar B_0}{Tg_F\mu_\mathrm{B}[b_\mathrm{q}^2\rho_0^2+ (1-2t/T)^2B_0^2]}\right\}^2},\label{eq:cdfield2}
\end{equation}
and,
\begin{align}B_0^\mathrm{CD}(t)&=B_0\frac{(1-2t)}{T} \nonumber\\
&+ \frac{8\hbar^2B_0^3(2t/T-1)}{g_F^2\mu_\mathrm{B}^2T^2[b_\mathrm{q}^2\rho_0^2+(1-2t/T)^2B_0^2]^2+4B_0^2\hbar^2}.\label{eq:cdfield3}
\end{align}

\begin{figure}[t!]
\centering
\includegraphics[width=0.45\textwidth]{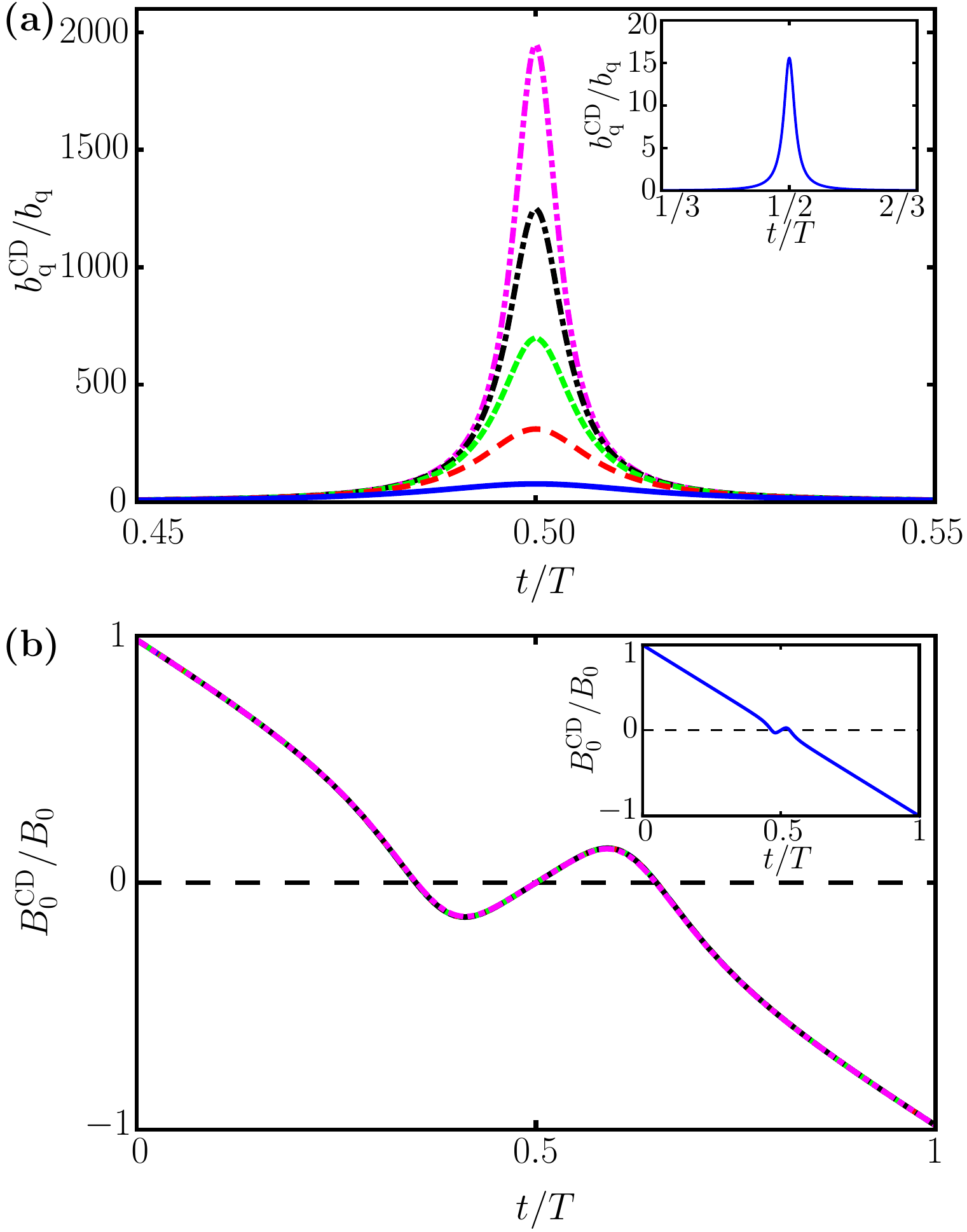}
\caption{(Color online) (a) Quadrupole magnetic field strength $b_\mathrm{q}^\mathrm{CD}$ and (b) bias magnetic field $B_0^\mathrm{CD}$ employed in the knot creation protocol as functions of time. Here, we set $T=0.1~\mathrm{ms}$, $B_0=50~\mathrm{mG}$, $b_\mathrm{q}=4.3~\mathrm{G/cm}$,  and $\rho_0=R/2$ (solid blue line), $R/4$ (dashed red), $R/6$ (dotted green), $R/8$ (dash-dotted black), and $R/10$ (dash-dot-dotted magenta), with $R=8.0~\mathrm{\mu m}$. These parameter values match those used in Sec.~\ref{sec:results2}. In the insets, the parameters are $T=1~\mathrm{ms}$, $B_0=0.1~\mathrm{G}$, and $\rho_0=R/2$, corresponding to a single knot. In (b), all the lines practically overlap.}\label{fig:bfields}
\end{figure}

%\onecolumngrid
\begin{figure*}[t]
\centering
\includegraphics[width=\textwidth]{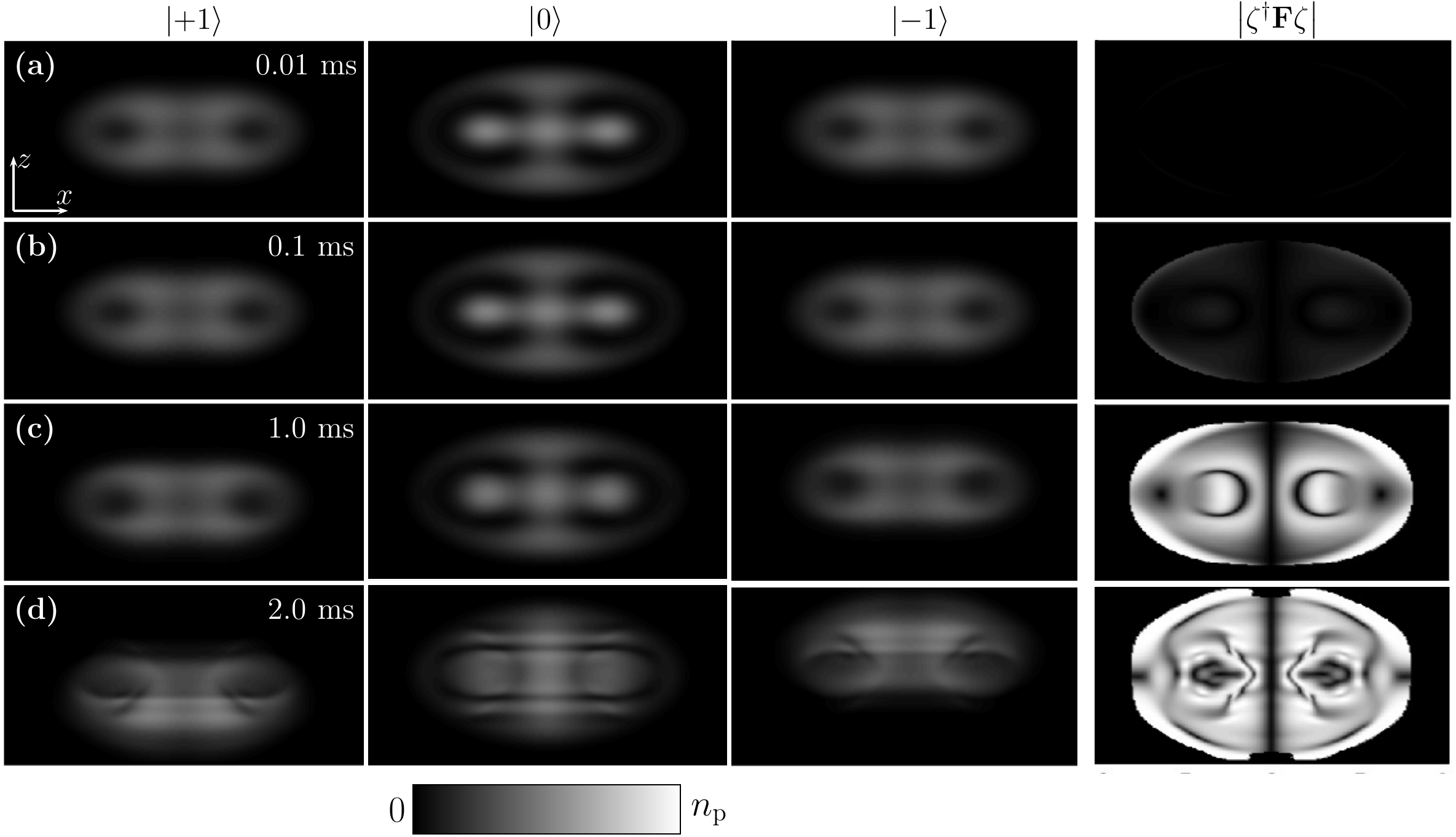}
\caption{Particle densities integrated along $y$ in different spin states and the spin density in a quantum knot for the creation time (a) $T=0.01~\mathrm{ms}$, (b) $0.1~\mathrm{ms}$, (c) 1.0$~\mathrm{ms}$, and (d) $2.0~\mathrm{ms}$. The first, second, and third columns correspond to spinor components $\zeta_{+1}$, $\zeta_0$, and $\zeta_{-1}$, respectively, and the fourth column corresponds to the spin density $\left|\zeta^\dagger{\bf F}\zeta\right|$ at the $y=0$ plane. Here, $\rho_0=R/2$, $B_0=0.5~\mathrm{G}$, the field of view in each panel is $20\times10~\mathrm{\mu m}^2$, and the peak particle density corresponds to $n_\mathrm{p}=2.5\times10^{11}~\mathrm{cm^{-2}}$. The peak spin density of all panels is normalized to unity.}\label{fig:partdens}
\end{figure*}
%\twocolumngrid

The control scheme of the magnetic field is presented in Fig.~\ref{fig:bfields}. In contrast to the control protocols used in Refs.~\cite{Kawaguchi:2008,Hall:2016}, the magnetic field zero point is not required to be centered in the middle of the condensate during the knot creation process, which is one of the most challenging experimental tasks~\cite{Ray:2014}. At the end of our creation protocol, the magnetic field zero point is naturally located far away from the condensate, whereas in Refs.~\cite{Kawaguchi:2008,Hall:2016} an additional control sequence is needed to achieve this condition. As we show below, by varying the parameter $\rho_0$ to a smaller value, our method allows for a convenient creation of knots with higher Hopf charge than that reported in Refs.~\cite{Kawaguchi:2008,Hall:2016}.

\section{\label{sec:results}Results}

We study the creation of quantum knots in the spin-1 BEC by numerically integrating the Gross--Pitaevskii equation~(\ref{eq:hamiltonian}) in the presence of the external magnetic field provided by the CD scheme as described by Eqs.~(\ref{eq:cdfield1})--(\ref{eq:cdfield3}). In the simulations, we employ a numerical grid of size $200^3$ with the typical volume $20\times20\times10~a_\rho^3$, accounting for the oblate shape of the condensate. Here, the harmonic oscillator length is identified as $a_\rho=\sqrt{\hbar/(\omega_\rho m)}= 1.0~\mathrm{\mu m}$. The effective extents of the ellipsoidal condensate are $R=8.0~\mathrm{\mu m}$ and $R_z=4.0~\mathrm{\mu m}$, chosen such that $\left|\Psi\right|^2<10^{-5}a_\rho^{-3}\approx 10^3~\mathrm{cm}^{-3}$ outside the ellipsoidal region. Throughout, we set $b_\mathrm{q}=4.3~\mathrm{G/cm}$ and for the simulations in Sec.~\ref{sec:results1} (Sec.~\ref{sec:results2}) we set $B_0=0.5~\mathrm{G}$ ($50~\mathrm{mG}$), such that $b_\mathrm{q}R\ll B_0$ is satisfied. The condensate is initially in the polar internal state $\hat{\bf z}=(0,1,0)^T_\mathrm{Z}$.

\subsection{\label{sec:results1}Creation of single knots}

Figure~\ref{fig:partdens} shows the $y$-integrated particle density distributions of different spinor components for various knot creation times. Here, we choose $\rho_0=R/2$ corresponding to a knot with the Hopf charge $Q=1$. For $T\le 1.0~\mathrm{ms}$ we numerically confirm the Hopf charge to be unity. The componentwise densities are also consistent with the knot structure: $\zeta_0$ component, corresponding to $\hat{\bf d}$ pointing to positive or negative $z$ [see Eq.~(\ref{eq:polarop})], fills the central region and the boundary, as well as the core around the central axis of the condensate. The combination of $\zeta_{\pm1}$ components, corresponding to $\hat{\bf d}$ residing along the $xy$-plane, fills the toroidal volume in between the $\zeta_0$ component~\cite{Hall:2016}. %The componentwise densities are also consistent with the knot structure, with $\zeta_0$ component ($\hat{\bf d}$ pointing to positive or negative $z$) filling the central region and boundary, as well as the core around the central axis of the condensate. The $\zeta_{\pm1}$ components ($\hat{\bf d}$ residing along the $xy$-plane) fill the toroidal volume in between the $\zeta_0$ component~\cite{Hall:2016}. 

\begin{figure}[t]
\centering
\includegraphics[width=0.5\textwidth]{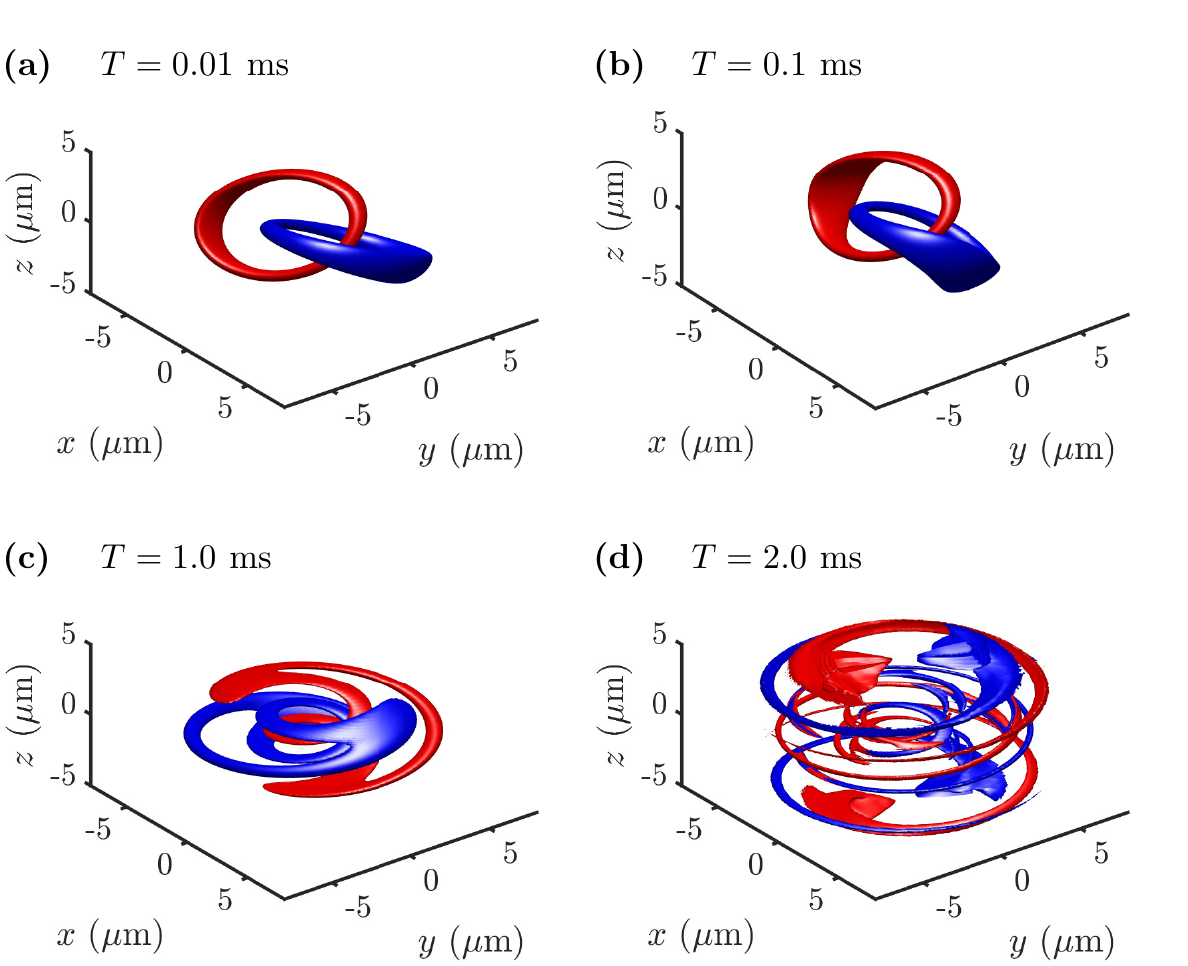}
\caption{(Color online) Preimages of nematic vectors $\hat{\bf d}=\hat{\bf x}$ (red region) and $\hat{\bf d}=-\hat{\bf x}$ (blue region) for (a) $T=0.01~\mathrm{ms}$, (b) $0.1~\mathrm{ms}$, (c) 1.0$~\mathrm{ms}$, and (d) $2.0~\mathrm{ms}$. Here, $\rho_0=R/2$, $B_0=0.5~\mathrm{G}$, and the surfaces show the volumes, inside which $|d_x|>0.97$.}\label{fig:rings}
\end{figure}

Ideally, the spin density vanishes for the polar phase. However, the spin density increases with the knot creation time, indicating a transition from the polar state to the ferromagnetic state in the condensate. We attribute the destruction of the knot structure at long creation times to this transition. The rapid decay of the polar phase is due to the spatial variations in the nematic vector field leading to spin currents~\cite{Kawaguchi:2008}. The transition to the ferromagnetic phase is further evidenced by the spatially separated $\zeta_{\pm1}$ states for $T=2.0~\mathrm{ms}$. 

The calculated preimages of $\hat{\bf d}=\hat{\bf x}$ and $\hat{\bf d}=-\hat{\bf x}$, shown in Fig.~\ref{fig:rings}, display two linked rings. The preimages are closed curves in real space, along which the nematic vector points to a constant direction. The linked structure starts to depart from the conventional Hopf link as the knot creation time increases. Finally, for $T>1.0~\mathrm{ms}$, the link cannot be identified and the Hopf charge vanishes. 

\subsection{\label{sec:results2}Creation of nested knots}

\begin{figure}[t!]
\centering
\includegraphics[width=0.5\textwidth]{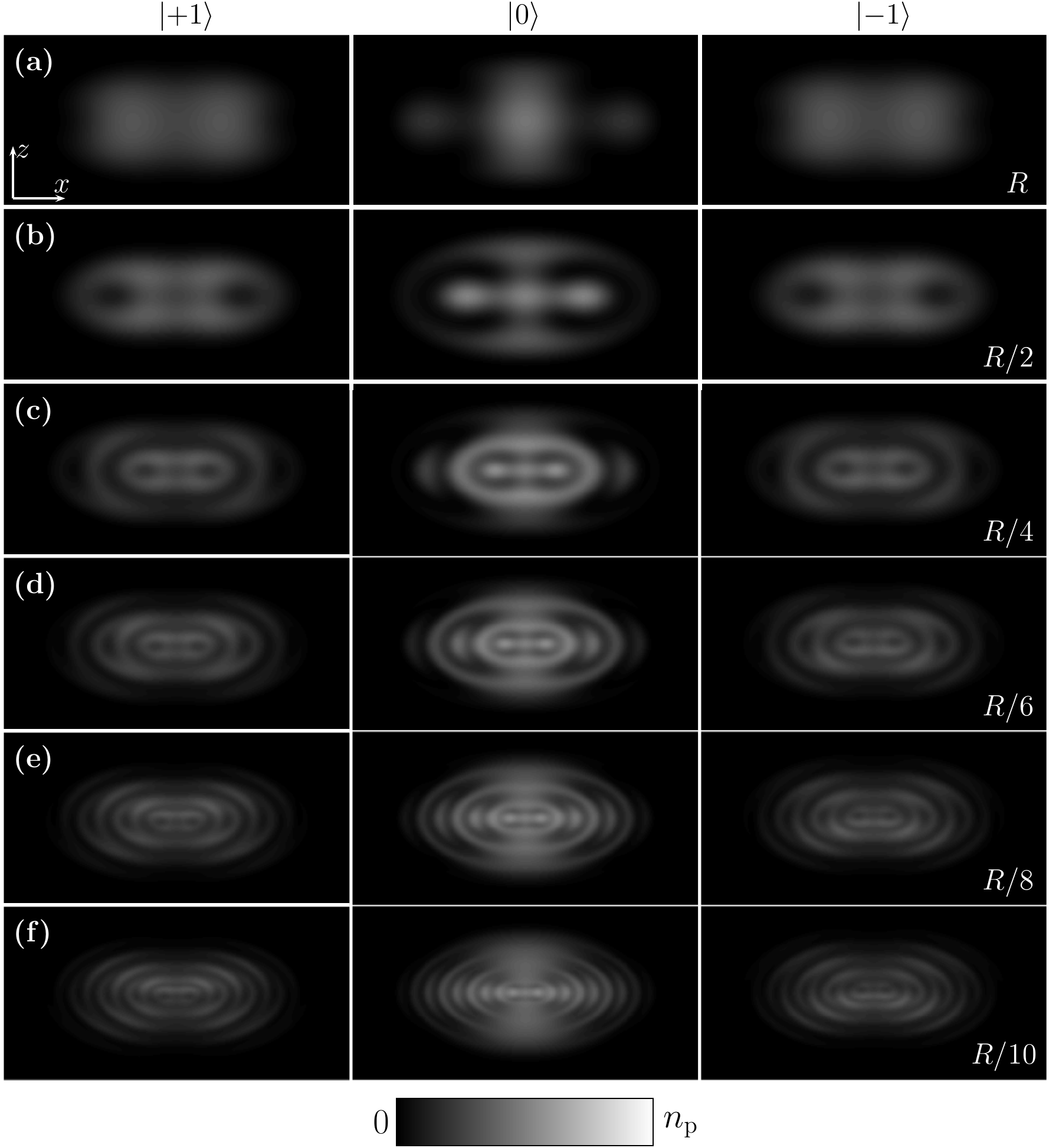}
\caption{Column particle densities of different spin states in a quantum knot as indicated for (a) $\rho_0=R$, (b) $R/2$, (c) $R/4$, (d) $R/6$, (e) $R/8$, and (f) $R/10$. Here, $T=0.1~\mathrm{ms}$, $B_0=50~\mathrm{mG}$, the field of view in each panel is $20\times10~\mathrm{\mu m}^2$, and the peak particle density is $n_\mathrm{p}=2.5\times10^{11}~\mathrm{cm^{-2}}$.}\label{fig:multiple}
\end{figure}

\begin{figure*}[t]
\centering
\includegraphics[width=0.85\textwidth]{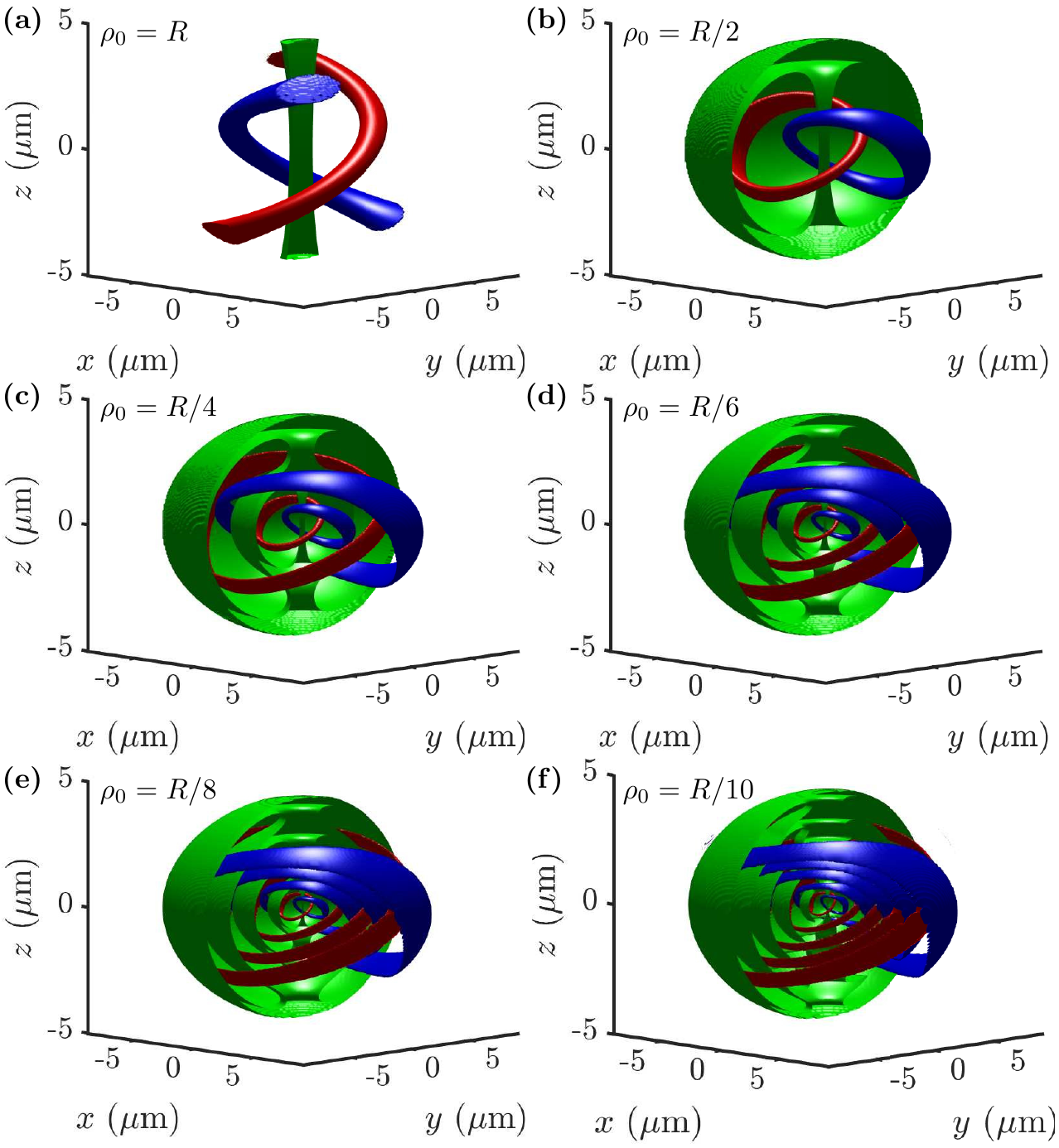}
\caption{(Color online) Preimages of $\hat{\bf d}=\hat{\bf x}$ (red region), $\hat{\bf d}=-\hat{\bf x}$ (blue region), and $\hat{\bf d}=\hat{\bf z}$ (green region) in a created quantum knot for (a) $\rho_0=R$, (b) $R/2$, (c) $R/4$, (d) $R/6$, (e) $R/8$, and (f) $R/10$. Here, $T=0.1~\mathrm{ms}$, $B_0=50~\mathrm{mG}$, and the shown surfaces enclose the volumes inside which $|d_x|>0.95$ or $d_z>0.95$. The green region is not shown for $x>0$.}\label{fig:ringsr0}
\end{figure*}

The particle densities and the calculated preimages for various choices of $\rho_0$ are shown in Figs.~\ref{fig:multiple} and \ref{fig:ringsr0}, respectively, with $T=0.1~\mathrm{ms}$ and $B_0=50~\mathrm{mG}$. The calculated Hopf charge increases with decreasing $\rho_0$ and the particle density distributions show the increase in the number of knot cores as $\rho_0$ decreases. The particle density distributions are consistent with those of multiple nested knot structures. The number of linked rings in the preimages increases according to the Hopf charge. %The Hopf charge follows the theoretical value $Q_\mathrm{ideal}(\rho_0) = R/(2\rho_0)$.

In the cases with Hopf charge $Q>1$, two linked rings appear $Q$ times in a nested structure, as is evident from the preimages in Fig.~\ref{fig:ringsr0}(c--f). These cases require a more careful topological inspection. Let us take $Q=2$ as an example and, for clarity, consider the scaled coordinate system $(x',y',z')=(x,y,2z)$ in which the condensate is spherical. The preimages of $\hat{\bf d}=\pm\hat{\bf x}'$ display two Hopf links. The two links are disconnected from each other, such that the inner link resides in the region $r'<R'/2$ and the outer link in $R'/2<r'<R'$. This holds for all choices of two different vectors $\hat{\bf d}\neq\hat{\bf z}'$.

The preimage of $\hat{\bf d}=\hat{\bf z}'$ includes a line along the $z'$-axis as well as two spheres with radii $R'$ and $R'/2$. The inner sphere with radius $R'/2$ can be compactified into a point, since $\hat{\bf d}=\hat{\bf z}'$ throughout the surface, thus compactifying the three-dimensional ball with $r'\le R'/2$ into $S^3$. This compactification procedure defines the usual Hopf map in the region $r'\le R'/2$. 

%\pagebreak
%\onecolumngrid

%\twocolumngrid

Topologically, the outer region is now homeomorphic to a three-dimensional ball with $r'\le R'$ as the sphere at $r'=R'/2$ is compactified into a point as described above. The outer sphere at $r'=R'$ is further compactified into another point, giving rise to another appearance of the Hopf map in the region $R'/2 \le r' \le R'$. Similar compactification procedures can be applied for the cases with $Q>2$, giving rise to the $Q$-fold nested Hopf maps.

\section{\label{sec:conclusion}Conclusion}

We have numerically studied an unusual application of the CD protocol to create topological knot structures in the nematic vector field of spin-1 BECs. Using this precise control scheme for the external magnetic field, knots with unit Hopf charge are created in the simulations for magnetic fields ramp times $10~\mathrm{\mu s}\le T \le1~\mathrm{ms}$. For longer ramp times the spin density is observed to increase in the condensate and the polar phase decays into the ferromagnetic phase, and consequently the knot structure is lost. Furthermore, our results show that knots with Hopf charge up to $Q=5$ can be created by varying the parameter $\rho_0$, which determines the radius of the core of knot. Knots with $Q>1$ exhibit interesting topology with nested Hopf links repeating $Q$ times.

\begin{acknowledgments}
We thank Yuki Kawaguchi, David Hall, and Konstantin Tiurev for discussions. 
We acknowledge funding 
by the Academy of Finland through its Centres of Excellence Program (Grants No. 251748 and No. 284621), 
by the European Research Council under Consolidator Grant No. 681311 (QUESS), 
by the KAUTE Foundation, and 
by Japan Society for the Promotion of Science (JSPS) Grants-in-Aid for Scientific Research (Grant No. 17K05554). 
This work is also supported by JSPS and Academy of Finland
Research Cooperative Program (Grant No. 308071).
CSC--IT Center for Science Ltd. (Project No. ay2090) and 
Aalto Science-IT project are acknowledged for computational resources. 
\end{acknowledgments}

\bibliography{cdknot}
\bibliographystyle{apsrev4-1}

\end{document}